\documentclass{PoS}

\usepackage{epsfig,multicol,multirow}
\usepackage{amsmath,subfigure,latexsym,amssymb}
\usepackage{wrapfig}

\usepackage{fontenc,times,mathptmx,graphicx}

\newcommand {\beq} {\begin{equation}}
\newcommand {\eeq} {\end{equation}}
\newcommand {\beqa}{\begin{eqnarray}}
\newcommand {\eeqa}{\end{eqnarray}}

\newcommand {\tr}{{\rm tr\,}}
\newcommand {\Tr}{\mbox{Tr\,}}

\title{Testing the AdS/CFT correspondence 
by Monte Carlo calculation of BPS and non-BPS Wilson loops 
in 4d $\mathcal{N}=4$ super Yang-Mills theory}

\ShortTitle{Testing AdS/CFT by Monte Carlo calculation of Wilson loops}

\author{\speaker{Masazumi Honda}$^{a}$, Goro Ishiki$^{b}$, 
Jun Nishimura$^{a,c}$ and Asato Tsuchiya$^{d}$
\vspace*{0.5cm} \\
\llap{$^a$}Department of Particle and Nuclear Physics,\\
Graduate University for Advanced Studies (SOKENDAI),\\
Tsukuba, Ibaraki 305-0801, Japan\\
\llap{$^b$}Department of Physics, Kyoto University,\\
3 Kyoto 606-8502, Japan\\
\llap{$^c$}KEK Theory Center, 
High Energy Accelerator Research Organization (KEK),\\
Tsukuba, Ibaraki 305-0801, Japan\\
\llap{$^d$}Department of Physics, Shizuoka University,\\
836 Ohya, Suruga-ku, Shizuoka 422-8529, Japan
\vspace*{0.5cm} \\
\email{mhonda@post.kek.jp},\email{ishiki@post.kek.jp},\\
\email{jnishi@post.kek.jp},\email{satsuch@ipc.shizuoka.ac.jp}}

\abstract{We test the AdS/CFT correspondence by calculating
Wilson loops in ${\cal N}=4$ super Yang-Mills theory on $R \times S^3$
in the planar limit.
Our method is based on a novel large-$N$ reduction, which
reduces the problem to
Monte Carlo calculations in the plane-wave matrix model
or the BMN matrix model, which is a 1d gauge theory
with 16 supercharges. By using the gauge-fixed momentum-space simulation,
we obtain results respecting 16 supersymmetries.
We report on the Monte Carlo results for the BPS circular Wilson loop, 
which reproduce the exact result up to strong coupling.
As a future prospect, we calculate a track-shaped Wilson loop
from the gravity side, which shows that a clear test of the AdS/CFT 
for the non-BPS case is also feasible.
}

\FullConference{The XXIX International Symposium on Lattice Field Theory, 
Lattice2011\\
July 10-16, 2011\\
Squaw Valley Lake Tahoe, California}

\begin{document}

\section{Introduction}

The gauge-gravity duality \cite{AdS-CFT} has been one of the
most important subjects in string theory over the past decade.
The most typical example is
the so-called AdS/CFT correspondence
between type IIB superstring theory on $AdS_5\times S^5$
and 4d ${\cal N}=4$ SU($N$) super Yang-Mills theory (SYM).
Even in this case, however,
a complete proof of the duality is still missing.
One of the reasons is that 
the parameter region described by the classical supergravity 
on the string theory side
corresponds to the strongly coupled region in the planar large-$N$ limit 
on the gauge theory side.
In order to study the strongly coupled 4d ${\cal N}=4$ SYM 
from first principles, 
one needs to have a non-perturbative formulation such as the lattice QCD.
The problem here is that the lattice regularization necessarily breaks 
translational symmetry, which is included in the supersymmetry (SUSY).
In order to restore SUSY in the continuum limit, 
one generally has to fine-tune parameters in the lattice 
action\footnote{Recently it was claimed
that fine-tuning can be avoided
at least up to the 1-loop level 
by using a lattice formulation with topological twist \cite{Catterall:2011pd}.}.

In this work
we adopt an alternative regularization method 
based on the idea of the large-$N$ reduction \cite{EK},
which preserves 16 supersymmetries.
Since 4d ${\cal N}=4$ SYM has conformal symmetry,
the theory on $R^4$ is equivalent to the theory on $R\times S^3$ 
through conformal mapping.
The novel large-$N$ reduction \cite{Ishii:2008ib} 
relates this theory to a reduced model,
which can be obtained by shrinking the $S^3$ to a point.
The resulting one-dimensional gauge theory with 16 supercharges can
be studied by using the gauge-fixed momentum-space 
simulation \cite{Hanada:2007ti} 
as in recent studies of the
D0-brane system \cite{BFSS_sim}.
Thus we can perform Monte Carlo calculations
in 4d ${\cal N}=4$ SYM respecting SUSY maximally and 
without fine-tuning\footnote{See refs.\ \cite{Hanada:2010kt}
for proposals for finite $N$.}.

We are going to test the AdS/CFT correspondence by calculating 
BPS and non-BPS Wilson loops in 
4d $\mathcal{N}=4$ SYM\footnote{See
refs.\ \cite{Nishimura:2009xm,lattice10} for some preliminary results on 
the Wilson loop and correlation functions.}. 
In particular, we reproduce an exact result for 
the circular Wilson loop, which serves as a check of our method.
For a non-trivial test of the AdS/CFT correspondence,
we consider the track-shaped Wilson loop
as an example of non-BPS operators, which 
cannot be calculated by analytic methods relying on SUSY.
We calculate it on the gravity side
by numerically solving a classical string equation of motion. 


\section{Large-$N$ reduction for ${\cal N}=4$ SYM on $R\times S^3$}
\label{sec:large-N-red}
Let us first discuss the novel large-$N$ reduction
for ${\cal N}=4$ SYM on $R\times S^3$.
By collapsing the $S^3$ to a point,
we obtain the plane wave matrix model (PWMM) 
or the BMN matrix model \cite{Berenstein:2002jq}\footnote{Properties 
of this model at finite temperature are studied at weak coupling
\cite{Kawahara:2006hs,eff_PWMM}
and at strong coupling \cite{Catterall:2010gf}.},
whose action is given by
\begin{eqnarray}
S_{\rm PW}
&=& \frac{1}{g_{\rm PW}^2}
\int
dt \, \tr 
\left[\frac{1}{2}(D_tX_M)^2-\frac{1}{4}[X_M,X_N]^2
+\frac{1}{2}\Psi^{\dagger} D_t \Psi
-\frac{1}{2}\Psi^{\dagger}\gamma_M[X_M,\Psi] \right.\nonumber\\
&~& \quad \quad \left.+\frac{\mu^2}{2}(X_i)^2
+\frac{\mu^2}{8}(X_a)^2 +i\mu\epsilon_{ijk}X_iX_jX_k
+i\frac{3\mu}{8}\Psi^{\dagger}\gamma_{123}\Psi \right] \ .
\label{pp-action}
\end{eqnarray}
Here the parameter $\mu$ is 
related to the radius of $S^3$ as $R_{S^3}=\frac{2}{\mu}$,
and the covariant derivative is defined by
$D_t=\partial_t-i[A, \ \cdot \ ]$,
where $A(t)$, as well as $X_M(t)$ and $\Psi (t)$, 
is an $N\times N$ hermitian matrix.
The range of indices is given by 
$1 \le M,N \le 9$, $1 \le i,j,k \le 3$ and $4 \le a \le 9$.
The model has the SU$(2|4)$ symmetry with 16 supercharges.

The PWMM possesses many discrete vacua 
representing multi fuzzy spheres, which are given explicitly by
\begin{equation}
X_i=\mu \bigoplus_{I=1}^{\nu}
\Bigl( L_i^{(n_I)}\otimes {\bf 1}_{k_I} \Bigr) \ 
\quad \mbox{with}\ \ \sum_{I=1}^{\nu}n_Ik_I=N \ ,
 \label{background}
\end{equation}
where $L_i^{(r)}$ are
the $r$-dimensional irreducible representation of the SU$(2)$ algebra 
$[L_i^{(r)},L_j^{(r)}]=i \, \epsilon_{ijk} \, L_k^{(r)}$.
These vacua preserve the SU$(2|4)$ symmetry, and are all degenerate.

In order to retrieve the planar ${\cal N}=4$ SYM on $R \times S^3$,
one has to pick up a particular background from (\ref{background}),
and consider the theory (\ref{pp-action}) around it.
Let us consider the vacuum defined by
\begin{equation}
k_I=k \ , \quad n_I=n+I-\frac{\nu+1}{2} \quad \quad
\mbox{for\ \ $I =1, \cdots , \nu$} \ ,
\label{our background}
\end{equation}
and take the large-$N$ limit in such a way that
\begin{eqnarray}
k\rightarrow \infty \ , \
\frac{n}{\nu} \rightarrow \infty \ , \ 
\nu\rightarrow\infty \ ,
\quad \mbox{with} \;\;
\lambda_{\rm PW} \equiv \frac{g_{\rm PW}^2 k}{n}
\; \; \mbox{fixed} \ .
\label{limit}
\end{eqnarray}
%
Then the resulting theory is claimed
\cite{Ishii:2008ib} to be equivalent\footnote{
See refs.\ \cite{earlier_novel} for earlier studies that led to 
this proposal. 
This equivalence was 
checked at finite temperature in the weak coupling regime \cite{eff_PWMM}.
It has also been extended to general group manifolds and 
coset spaces \cite{group-coset}.}
to the planar limit of ${\cal N}=4$ SYM on $R \times S^3$
with the 't Hooft coupling constant given by
\beq
\lambda_{\rm SYM} =2\pi^2 \lambda_{\rm PW}(R_{S^3})^3 
=\frac{16 \pi^2 k }{n}
\frac{g_{\rm PW}^2}{\mu^3} \ .
\label{lambda-sym}
\eeq
%

The above equivalence may be viewed as
an extension of the large-$N$ reduction \cite{EK},
which asserts that the large-$N$ gauge theories can be 
studied by dimensionally reduced models.
The original idea for theories compactified on a torus
can fail due to
the instability of the U(1)$^D$ symmetric vacuum
of the reduced model \cite{Bhanot}.
This problem is avoided in the novel proposal
since the PWMM is a massive theory and the vacuum preserves the maximal SUSY.
This regularization respects 16 supersymmetries, which is half of 
the full superconformal symmetry of ${\cal N}=4$ SYM on $R\times S^3$.
Since any kind of UV regularization breaks the conformal symmetry,
this regularization is optimal from the viewpoint of preserving SUSY.

\section{Wilson loops in 4d $\mathcal{N}=4$ SYM}
\label{sec:wloop}

Let us consider the following type of Wilson loop
\begin{equation}
W(C) = \frac{1}{N}\Tr \mathcal{P} 
   \exp{\oint_C ds \left( iA_\mu^{R^4} 
\dot{x}^\mu(s) +\left| \dot{x}^\mu (s) \right| X_a^{R^4} 
\theta_a  \right) }  \ ,
\end{equation}
where $\dot{x}^\mu (s)\equiv \frac{dx^\mu (s)}{ds}$ and $\theta_a$ 
is a constant which satisfies $\theta_a \theta_a =1$.
The fields $A_\mu^{R^4}$ and $X_a^{R^4}$ represent 
the gauge field and the six scalars in 
4d $\mathcal{N}=4$ SYM on $R^4$, respectively.
Due to the particular way in which the scalars appear,
one can obtain predictions from the gravity side based on 
the AdS/CFT correspondence as
\begin{equation}
\lim_{N\rightarrow\infty ,\lambda_{\rm SYM}\rightarrow\infty} 
\Big\langle W(C) \Big\rangle_{\rm SYM} 
=  e^{-S(C)} \ ,
\end{equation}
where $S(C)$ represents the area of the minimal surface spanning
the loop $C$ on the boundary of the AdS space.

For the circular Wilson loop $W(C_{\rm circ})$, 
which is a (1/2-)BPS operator,
there is an exact result on the gauge theory side,
which is obtained by summing up planar ladder diagrams 
or by using the localization method \cite{circular}.
The result is given by
\begin{eqnarray}
\lim_{N\rightarrow\infty} 
\Big\langle W(C_{\rm circ}) \Big\rangle_{\rm SYM} 
&=& \sqrt{\frac{2}{\lambda_{\rm SYM}}} \, 
I_1 \Big(\sqrt{2\lambda_{\rm SYM}} \Big) 
\label{localization} \\
&\simeq & \frac{e^{\sqrt{2\lambda_{\rm SYM}}}}
{\left( \frac{\pi}{2}\right)^{1/2}(2\lambda_{\rm SYM})^{3/4}} 
          \quad \quad  \rm{for}\ \lambda_{\rm SYM}\gg 1  \ ,
\label{strong}
\end{eqnarray}
where $I_1 (x)$ is the modified Bessel function of the first kind.
The result is independent on the radius of the circle,
which is a consequence of the scale invariance of $\mathcal{N}=4$ SYM. 
At strong coupling it agrees with the result obtained 
from the dual geometry 
$S(C_{\rm circ})=-\sqrt{2\lambda_{\rm SYM}}$ \cite{gravC}.
This is an explicit example of the AdS/CFT correspondence.
We use the exact result (\ref{localization}) for arbitrary $\lambda_{\rm SYM}$
to check our calculation method.

\begin{figure}[htbp]
\mbox{\raisebox{-45mm}{\includegraphics[width=7.2cm,clip]{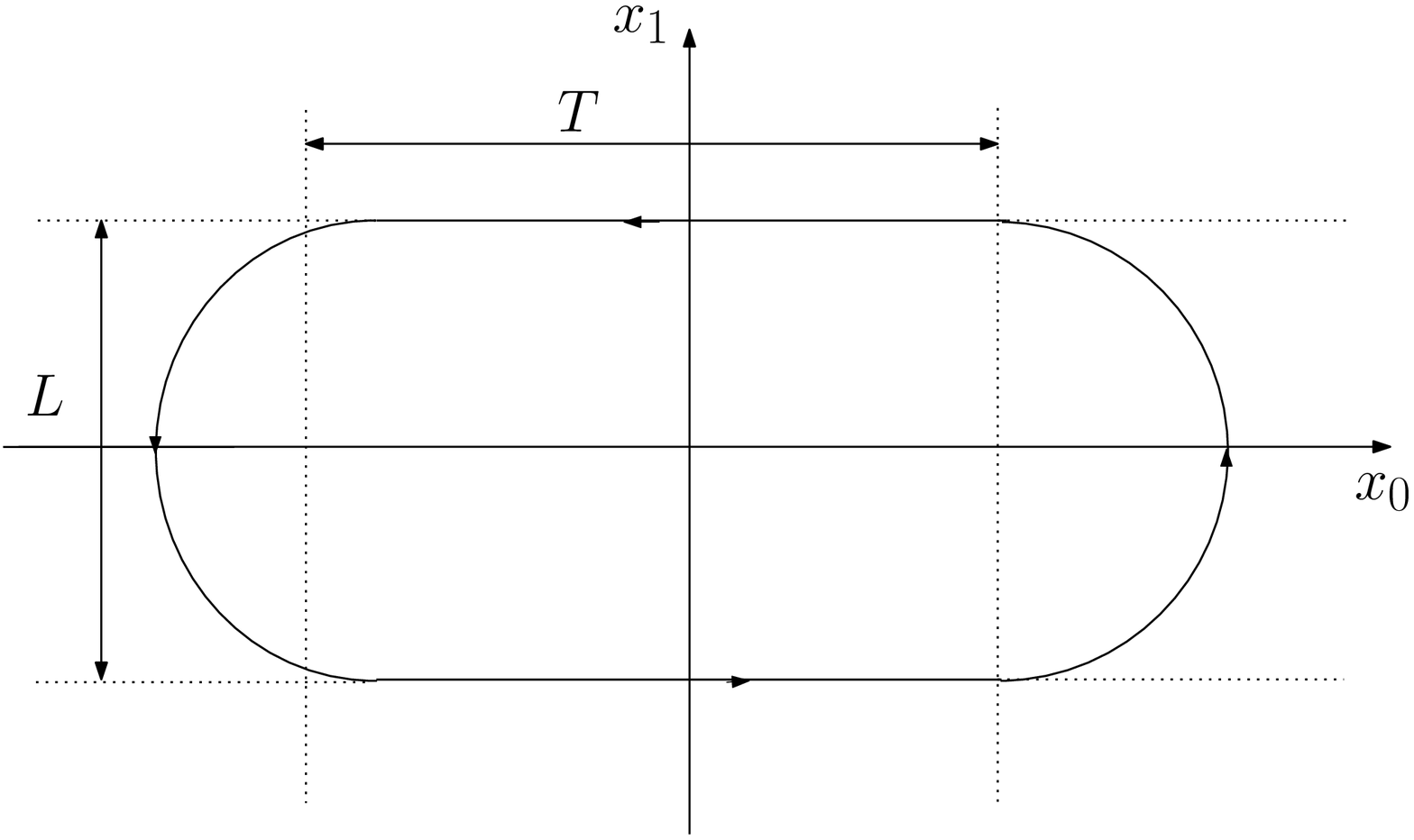}}}
\includegraphics[width=5.2cm,angle=270]{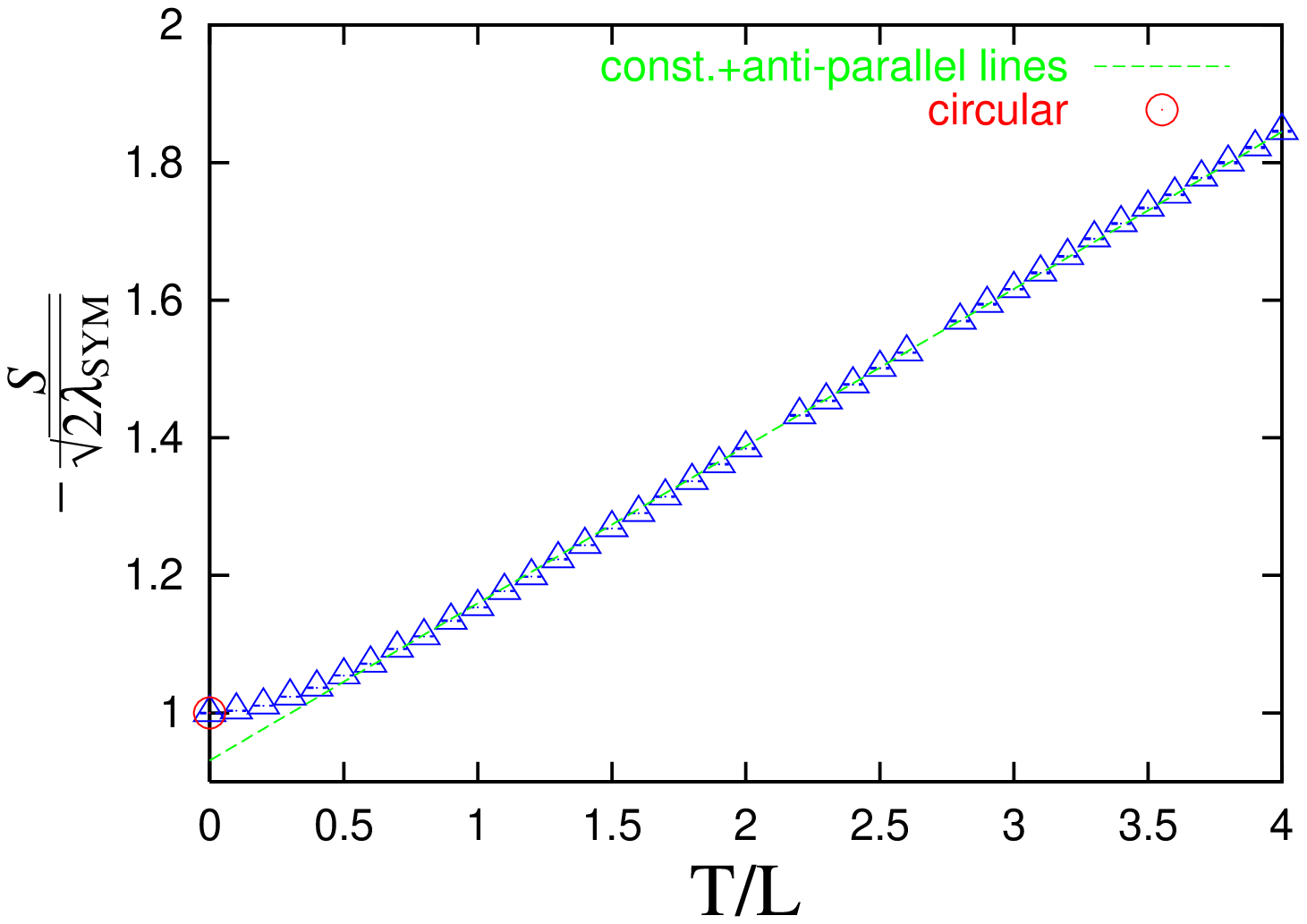}
 \caption{ 
(Left) The track-shaped Wilson loop
has two straight lines of length $T$, 
which are anti-parallel to each other,
and two semi-circles with radius $L/2$.
(Right) The numerical results for 
$-\frac{S( C_{{\rm track}(T , L)})}{\sqrt{2\lambda_{\rm SYM}}}$ 
is plotted as a function of $T/L$.
The circle at $T=0$ represents
the circular limit ($T\rightarrow 0$) and
the straight line represents a fit to 
the anti-parallel-line limit 
($T\rightarrow\infty$) with an additive constant treated 
as a free parameter.
}
  \label{fig:gravity}
\end{figure}

The track-shaped Wilson loop has two parameters $T$ and 
$L$ as depicted in figure \ref{fig:gravity} (Left).
It has two limits:
the $T\rightarrow 0$ limit,
which corresponds to the circular Wilson loop,
and $T\rightarrow\infty$, which corresponds to 
two anti-parallel lines.
Since the track-shaped Wilson loop for $T \neq 0$
is a non-BPS operator,
it cannot be calculated on the gauge theory side
by such analytic methods as the localization method 
which rely on SUSY.
The limit ($T\rightarrow\infty$) 
of two anti-parallel lines
is calculated based on the AdS/CFT correspondence
from the gravity side as \cite{rectangular}
\begin{equation}
\lim_{T\rightarrow\infty}\frac{1}{T} 
\ln
\Big\langle W(C_{{\rm track}(T , L)}) \Big\rangle_{\rm SYM} 
= -\sqrt{2\lambda_{\rm SYM}} 
\frac{4\pi^2}{\Gamma (1/4)^4}\frac{1}{L} \ ,
\end{equation}
which is consistent with the conformal symmetry of the theory,
and also with the fact that
$\mathcal{N}=4$ SYM on $R^4$ is in the Coulomb phase.
Here we extend the calculation on the gravity side 
to arbitrary $T/L$, and obtain explicit numerical results
by solving the classical string equation of motion.
In figure \ref{fig:gravity} (Right) we plot
the minimal surface obtained by Newton's method.

\section{Monte Carlo method}
\label{sec:monte}
In order to simulate the PWMM (\ref{pp-action}),
we compactify the $t$-direction to a circle of circumference $\beta$.
Since we are interested in the properties at zero temperature,
we impose periodic boundary conditions 
on both scalars $X_{M}(t)$ and fermions $\Psi_{\alpha}(t)$,
which keep SUSY intact.
In Fourier-mode simulation \cite{Hanada:2007ti},
we first fix the gauge symmetry completely by choosing
$A(t) = \frac{1}{\beta}{\rm diag} (\alpha_{1},\cdots ,\alpha_{N})$
with $-\pi <\alpha_{a}\leq \pi$,
and then make a Fourier expansion 
\ $X_{M}(t) = \sum_{n=-\Lambda}^{\Lambda} 
\tilde{X}_{M,n}e^{i\omega nt}\ (\omega\equiv \frac{2\pi}{\beta})$ 
and similarly for the fermions.
The upper bound $\Lambda$ on the Fourier modes plays 
the role of the UV cutoff.
The original PWMM can be retrieved by just taking 
the limits $\beta\rightarrow\infty$
and $\frac{\Lambda}{\beta} \rightarrow \infty$
since there are neither UV nor IR divergences.
The model regularized by finite $\beta$ and $\Lambda$ 
can be simulated by the RHMC algorithm.
This method has been applied extensively
to the D0-brane system corresponding to $\mu=0$,
and the results confirmed the gauge/gravity duality 
for various observables \cite{BFSS_sim}.\footnote{See
refs.\ \cite{Catterall-Wiseman} for Monte Carlo calculations
based on the lattice regularization.
}
Since the parameter $\mu$ in the action (\ref{pp-action})
can be scaled away by appropriate redefinition of fields and parameters,
we take $\mu =2\ (R_{S^3}=1)$ without loss of generality.

The Wilson loop in $\mathcal{N}=4$ SYM 
can be calculated in PWMM in the following way.
When we perform the conformal mapping from $R^4$ to $R\times S^3$,
the radial and angular directions are mapped to 
the time and $S^3$-directions, respectively.
Therefore, an arbitrary loop on a plane in $R^4$ is mapped to a loop
on $R\times S^3$, which can be projected to a great circle on $S^3$.
Such a Wilson loop can be represented in the large-$N$ reduced model as
\begin{equation}
W_{\rm red}(C) = \frac{1}{N}\Tr \mathcal{P} 
   \exp{\oint_C ds \left( i A_0 \frac{dt}{ds}  +i X_i e_\mu^i  \dot{x}^\mu (s) 
        +\left| \dot{x}_\mu (s) \right| X_a \theta_a  \right) }  \ ,
\label{op_red}
\end{equation}
where $e_j^i = e_j^i (x^\mu (s))$ is the dreibein on $S^3$.
The expectation value of this operator is related 
to the average of the original Wilson loop as \cite{Ishii:2007sy}
\begin{equation}
\Big\langle W(C) \Big\rangle_{\rm SYM}
= \Big\langle W_{\rm red}(C) \Big\rangle \ ,
\label{relation-op}
\end{equation}
where $\langle \cdots \rangle$ on the right-hand side denotes the expectation value 
in the large-$N$ reduced model (PWMM).
In the case of circular Wilson loop, the relation (\ref{relation-op})
was confirmed by reproducing the SYM result (\ref{localization})
from the reduced model to all orders in perturbation theory
assuming that non-ladder diagrams do not contribute \cite{Ishiki:2011ct}.

\section{Numerical results}
\label{sec:result}
\begin{figure}[htbp]
  \begin{center}
   \includegraphics[width=7.2cm]{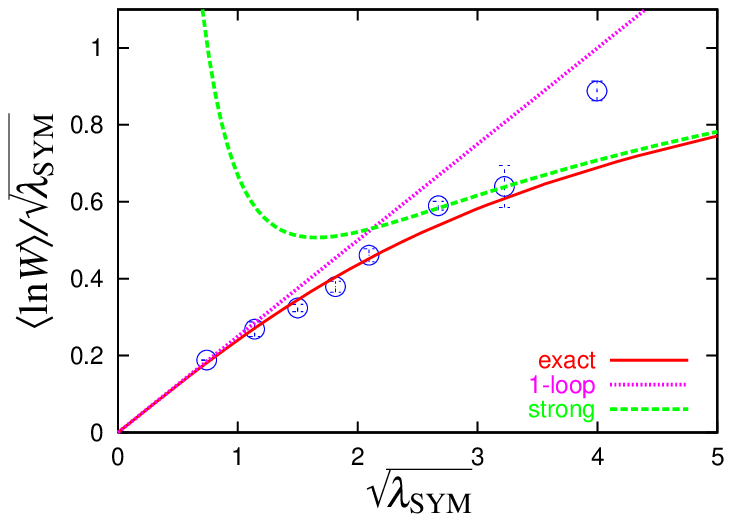}
\begin{center}	
\begin{tabular}[t]{|c||c|c|c|c|c|c|c|c|}
\hline
$g_{\rm PW}^2 N$     & 0.13  & 0.30 & 0.51 & 0.75 & 
1.00 & 1.63 & 2.37 & 3.64  \\ \hline 
$\beta$              & 10.0  & 7.50 & 6.25 & 
5.50 & 5.00 & 4.25 & 3.75 & 3.25     \\ 
\hline \hline
$\lambda_{\rm SYM}$  & 0.55  & 1.30 & 2.25 & 
3.30 & 4.39 & 7.14 & 10.4 & 16.0  \\ 
\hline
\end{tabular}
\end{center}
  \end{center}
  \caption{The log of the circular Wilson loop normalized 
by $\sqrt{\lambda_{\rm SYM}}$ is plotted against $\sqrt{\lambda_{\rm SYM}}$.
The values of $g_{\rm PW}^2 N$ and $\beta$ 
(and the corresponding values of $\lambda_{\rm SYM}$) 
we use are listed in the table.
The solid line represents the exact result (\protect\ref{localization}). 
The dashed line represents the behavior (\protect\ref{strong})
at strong coupling, whereas the dotted line represents 
the leading perturbative behavior 
$\ln{\langle W\rangle}\simeq \frac{1}{4}\lambda_{\rm SYM}$.
}
\label{fig:circle}
\end{figure}

In figure \ref{fig:circle} we present our preliminary results 
for the circular Wilson loop.
We have performed the $\Lambda\rightarrow\infty$ extrapolation
using $\Lambda$ = 6,8,10,12
assuming that finite $\Lambda$ effects are O($1/\Lambda$). 
The background is chosen to be $(n ,\nu )=(\frac{3}{2},2 )$ and
we performed an extrapolation to $k=\infty$ using the data 
for $k = 2,3,4,5$ assuming that the finite-k effects are
O($1/k^2$).
We also plot the exact result (\ref{localization}).
Except for the data point at $\sqrt{\lambda_{\rm SYM}}=4$,
the agreement with the exact result is promising.
Note, in particular, that we already
start to observe a bent from the weak coupling behavior 
towards the strong coupling behavior.
This is remarkable considering the rather small matrix size.
We consider this as a result of the fact that our formulation
respects sixteen supersymmetries.

\section{Summary and discussions }
\label{sec:summary}

We 
have investigated
nonperturbative properties of the 4d SU($\infty$) $\mathcal{N}=4$ SYM
from first-principles respecting 16 supersymmetries.
In particular, we have reproduced the exact result 
(\ref{localization}) up to $\lambda_{\rm SYM}\simeq 10.4$.
As a nontrivial check of the AdS/CFT correspondence,
we are planning to study the track-shaped Wilson loop.
We have obtained explicit results on the gravity side by 
numerically solving a classical string equation of motion.
The results nicely interpolate the two limits, which 
are already calculated, i.e., 
the circular loop and the anti-parallel lines.
We hope to report on the calculations on the gauge theory side
in the forth-coming publication.


\acknowledgments
We thank H.\ Kawai and Y.\ Kitazawa for valuable discussions.
Computation was carried out on PC clusters at
KEK and Fermilab.
The work of M.\ H.\ is supported by 
Japan Society for the Promotion of Science (JSPS).
The work of G. I. is supported by the Grant-in-Aid for the Global COE 
Program "The Next Generation of Physics, Spun from Universality and 
Emergence" from the Ministry of Education, Culture, Sports, Science 
and Technology (MEXT) of Japan.
The work of J.\ N.\ and A.\ T.\ is supported
by Grant-in-Aid for Scientific
Research (No.\ 19340066, 20540286, 19540294 and 23244057)
from JSPS.


\end{document}